\documentclass[10pt]{article}

\usepackage{amsmath}
\usepackage{extsizes}
\usepackage{amsfonts}
\usepackage{amssymb}
\usepackage{mathrsfs}
\usepackage{graphicx}
\usepackage{verbatim}
\usepackage[stable]{footmisc}
\usepackage{epstopdf}
\usepackage{bm}
\usepackage[colorlinks,linkcolor=blue,citecolor=blue,urlcolor=blue]{hyperref}
\usepackage{tabularx} 
\usepackage{floatrow}
\usepackage{lipsum}
\usepackage{color}
\usepackage[usenames,dvipsnames]{xcolor}
\usepackage{soul}

\begin{document}
%\frontmatter
	\title{Vibronic spectroscopy of sympathetically cooled CaH$^{+}$}
	\date{\today}
	\author{Ren\'{e} Rugango \thanks{School of Chemistry and Biochemistry, Georgia Institute of Technology, Atlanta, GA  30332,USA}\and Aaron T. Calvin \footnotemark[1] \and Smitha Janardan \footnotemark[1] \and Gang Shu \footnotemark[1] \and Kenneth R. Brown \footnotemark[1] \thanks{Schools of Computational Science and Engineering and Physics, Georgia Institute of Technology, Atlanta, GA 30332, USA} \thanks{\href{mailto:ken.brown@chemistry.gatech.edu}{ken.brown@chemistry.gatech.edu}}}

\evensidemargin=0.5in
\oddsidemargin=0.5in
\maketitle

\begin{abstract}
We report the measurement of the 1$^{1}\Sigma\longrightarrow$ 2$^{1}\Sigma$ transition of CaH$^+$ by resonance-enhanced photodissociation of CaH$^+$ that is co-trapped with laser-cooled Ca$^+$ . We observe four resonances that we assign to transitions from the vibrational $v$=0 ground state to the $v'$=1-4 excited states based on theoretical predictions.  A simple theoretical model that assumes instantaneous dissociation after resonant excitation yield results in good agreement with the observed spectral features except for the unobserved $v'$=0 peak.  The resolution of our experiment is limited by the mode-locked excitation laser, but this survey spectroscopy enables future rotationally resolved studies with applications in astrochemistry and precision measurement.\footnote{This is the pre-peer reviewed version of the following article: R.Rugango et al. ChemPhysChem, online, (2016), which has been published in final form at \href{http://onlinelibrary.wiley.com/doi/10.1002/cphc.201600645/abstract}{ DOI: 10.1002/cphc.201600645}. This article may be used for non-commercial purposes in accordance with Wiley Terms and Conditions for Self-Archiving.}
\end{abstract}

\section{Introduction}
The spectroscopy of cold trapped ions is an important tool for the determination of molecular species in the interstellar medium \cite{sandra2014, maier2015} and the precise measurement of fundamental physical constants \cite{KajitaJPhysB2011,BresselPRL2012, LohScience2013}. Application of methods from the laser-cooling of atomic ions has led to new advances in taming both external and internal quantum states of trapped molecular ions. The new techniques have allowed for the sympathetic sideband cooling of motional modes to the ground state \cite{rugango2015, wan2015}, preparation of the rotational ground state \cite{LienNatComm2014, StaanumNatPhys2010,SchneiderNatPhys2010}, and the non-destructive probing of rotational states \cite{wolf2015}. The large rotational constants of diatomic hydrides \cite{StaanumNatPhys2010,SchneiderNatPhys2010,LienNatComm2014,HansenNature2014} which slows blackbody rethermalization after rotational cooling combined with the availability of atomic coolants of similar masses make these molecular ions particularly well-suited to high precision spectroscopy measurements. Homonuclear diatomic molecular ions allow almost complete decoupling from blackbody radiation and are also a promising route towards tests of fundamental constants \cite{Schiller2014, Germann2014, KajitaPRA2014}.

The detection of  neutral CaH radicals  in sun spots \cite{albert1909} suggests possible solar presence of CaH$^{+}$. However, the spectroscopy  vital to the detection of this molecular ion has been absent in contrast to other neutral metal hydrides. Short storage times and weak absorption and dispersion signals \cite{mcall2011} pose intrinsic challenges to molecular beam spectroscopy, a limitation that has contributed to the lack of experimental data on CaH$^{+}$. Within the last few decades, new trapping and laser-cooling techniques have made it possible to generate metal hydride ions with quantum state-controlled chemical reactions between laser-cooled trapped atomic ions and molecular hydrogen \cite {kimura, schiller2006, MolhavePRA2000}. These techniques have created a new opportunity for the spectroscopy of CaH$^{+}$  with the observation of a dissociative electronic transition \cite{HansenAngew2012}  and ground state vibrational overtones \cite{ncamiso2015}.

Apart from being of astronomical interest, CaH$^{+}$ has much potential as a candidate for testing the time variation of fundamental physical constants. Vibrational and rotational transitions in molecules have different dependencies on  the electron-to-proton mass ratio, $\sqrt{m_{e}/m_{p}}$ and $m_{e}/m_{p}$, respectively \cite{Flohlich2004, Roth2005, krems}. By comparing these transitions within one system, many systematic errors can be canceled. It has been previously proposed that the uncertainty in rovibrational frequency measurement in CaH$^{+}$ is on the order of the predicted astronomical time variation of $m_{e}/m_{p}$ \cite{kajita2012}.

Despite the lack of experimental  data, CaH$^{+}$ spectroscopy has been the subject of many previous theoretical studies. \textit{Ab initio} calcuations of vibrational dipole moments of metal hydrides \cite{AbeKajita} by Abe \textit{et al.}, and subsequent calcuations of potential energy curves of the electronic ground and excited states of the CaH$^{+}$ \cite{KajitaJPhysB2011, olivier2012} have offered theoretical guidance  for our experiments. We report our measurement of the 1$^{1}\Sigma\longrightarrow$ 2$^{1}\Sigma$ vibronic transitions by resonance enhanced dissociation between 370 nm and 421 nm.

\section{Experiment}
\begin{figure*}
	\centering
	\includegraphics[width=13cm]{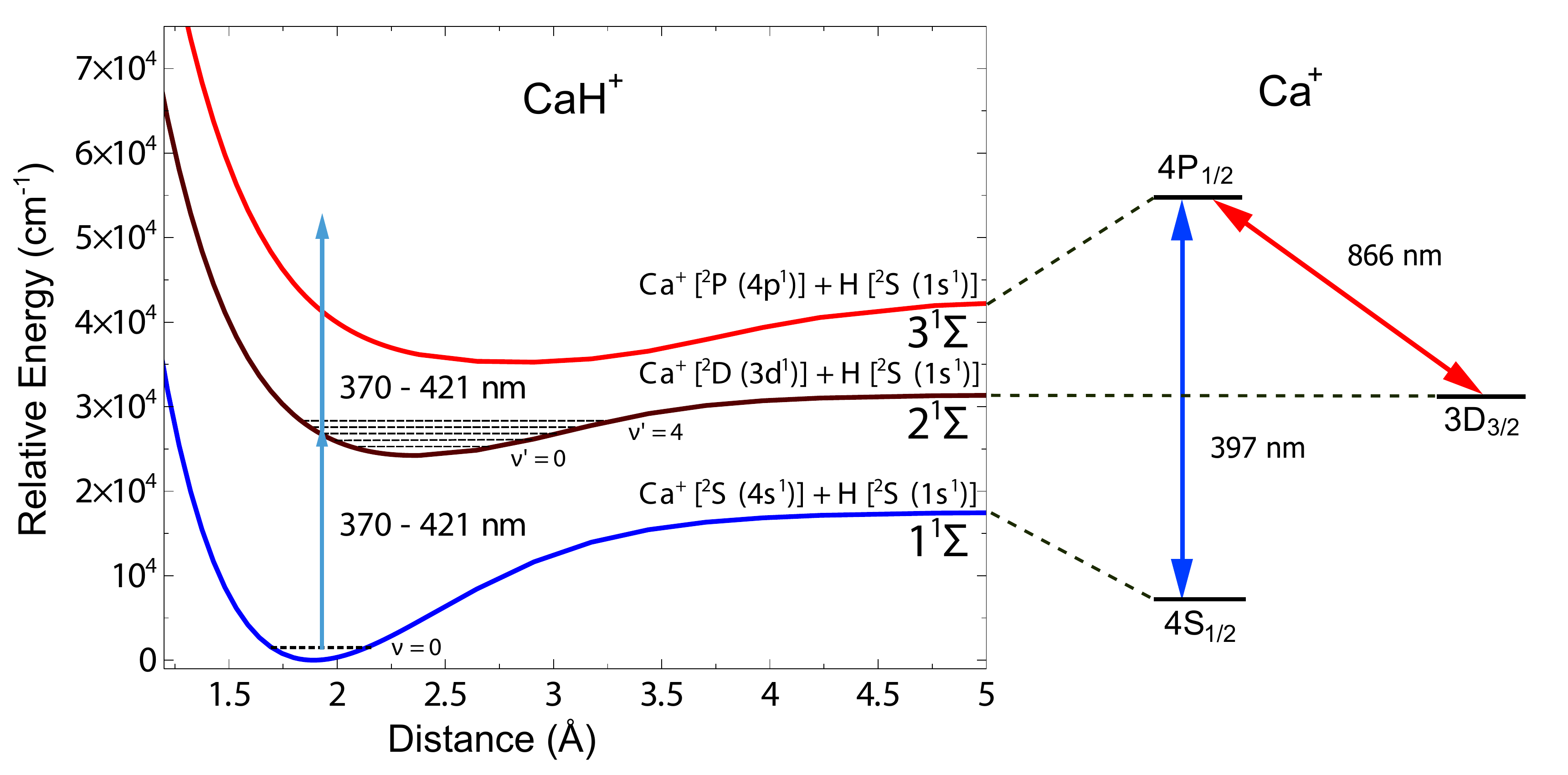}
	\caption{The three lowest $^1\Sigma$ potential energy curves for CaH$^{+}$  are shown \cite{KajitaJPhysB2011}. We measure vibrational lines using resonance enhanced dissociation. The first photon excites the molecules on the 1$^{1}\Sigma\longrightarrow$ 2$^{1}\Sigma$ transition. A second photon could then dissociate the ion.   The atomic asymptotes of the plotted molecular potentials lead to Ca$^+$ in different electronic states and H in the ground state. A simplified Ca$^+$ energy level diagram shows the states involved in Doppler cooling and how they correspond to the limits of the molecular states (energy not to scale) . The 397 nm laser is the main Doppler cooling laser and the 866 nm laser repumps the metastable D$_{3/2}$ state.} 
	\label{PES}
\end{figure*}

The ion trap used in this experiment was previously described in Ref. \cite{goeders}. The experiment takes place in a spherical octagon vacuum chamber (Kimball Physics MCF800-SphOct-G2C8) with eight viewports used for electrical connections and optical access. Low DC voltages (0 - 10V) applied to all eleven pairs of DC electrodes weakly confine the ions in the axial direction, and a RF voltage of 210 V oscillating at 19.35 MHz traps  ions radially.  397 nm and 866 nm lasers Doppler cool the ions axially and radially. Figure \ref{PES} outlines the transitions used for Doppler cooling Ca$^+$ ions.

Photoionization of sublimed neutral Ca allows the trapping of a few hundred Ca$^{+}$ ions. Addition of H$_{2}$ gas to the vacuum chamber through a manual leak valve promotes the formation of CaH$^{+}$ through photo-activated reactions between Ca$^{+}$ and H$_{2}$. Upon addition of hydrogen gas, the pressure in the chamber goes from 1.2$\times$10$^{-10}$ Torr to around 3$\times$10$^{-8}$ Torr. At these pressures, decrease in fluorescence on the photomultiplier tube (PMT) (Hamamatsu H7360-02) accompanied by the darkening of a large part of the Coulomb crystal on the CCD camera (Princeton Instruments Cascade 1K) indicates 50 to 100 reactions within a few minutes. The molecular ions are excited using a frequency-doubled Ti:sapphire laser. We achieve a 10$\%$ efficiency in laser power after doubling the IR beam through a 0.1 mm thick Foctek BBO crystal. The doubled beam enters along the trap axis, and dissociation of the molecular ions is detected by the increase in fluorescence, as well as the reappearing of the bright Ca$^{+}$ ions .

The frequency-doubled Ti:sapphire is tuned between 370 and 421 nm.  Theoretical calculations predicted \cite{KajitaJPhysB2011, olivier2012} that the 1$^{1}\Sigma\longrightarrow$ 2$^{1}\Sigma$ transition would occur in this range, Figure  \ref{PES}.  
The bandwidth of pulses is measured for different wavelengths by an Ocean Optics spectrometer (model HR2000+) and corresponds to a pulse length of 300 fs. The power is maintained at 20 mW for all wavelengths.

\begin{figure*}
	\centering
	\includegraphics[width=11.2cm]{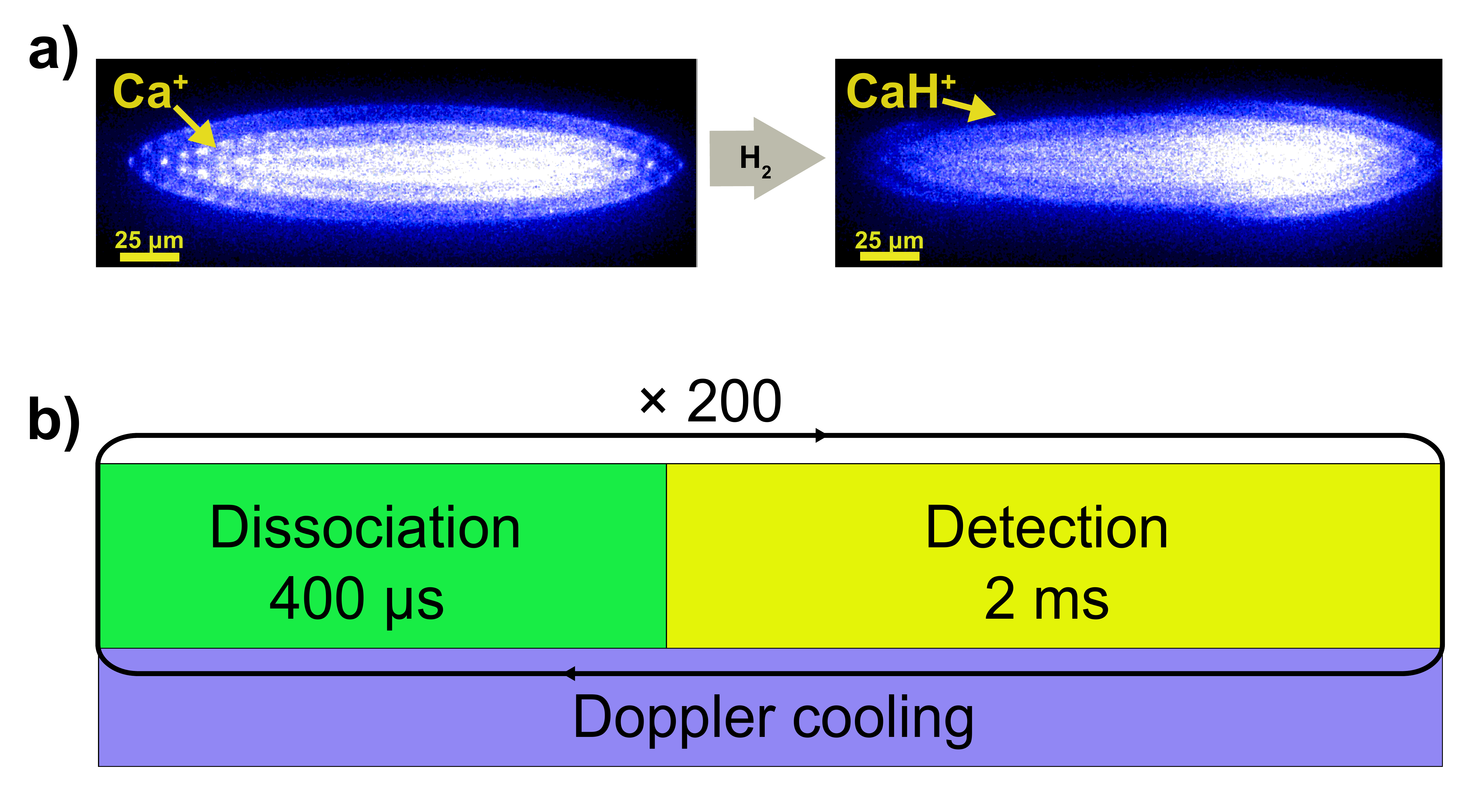}
	\caption{a) Coulomb crystal containing a few hundred Ca$^{+}$ ions react with H$_{2}$ to make 50 to 100 CaH$^{+}$ ions. The reaction is indicated by the darkening of the left side of the crystal. The asymmetry seen in the Coulomb crystal is due to the 397 nm radiation pressure on Ca$^{+}$ ions. b) The pulse sequence used to dissociate the molecular ions is shown. The Doppler cooling lasers remain on to avoid ion heating and losses. An AOM switches the dissociation beam.}
	\label{reaction}
\end{figure*}

\begin{figure*}
	\centering
	\includegraphics[width=11.2cm]{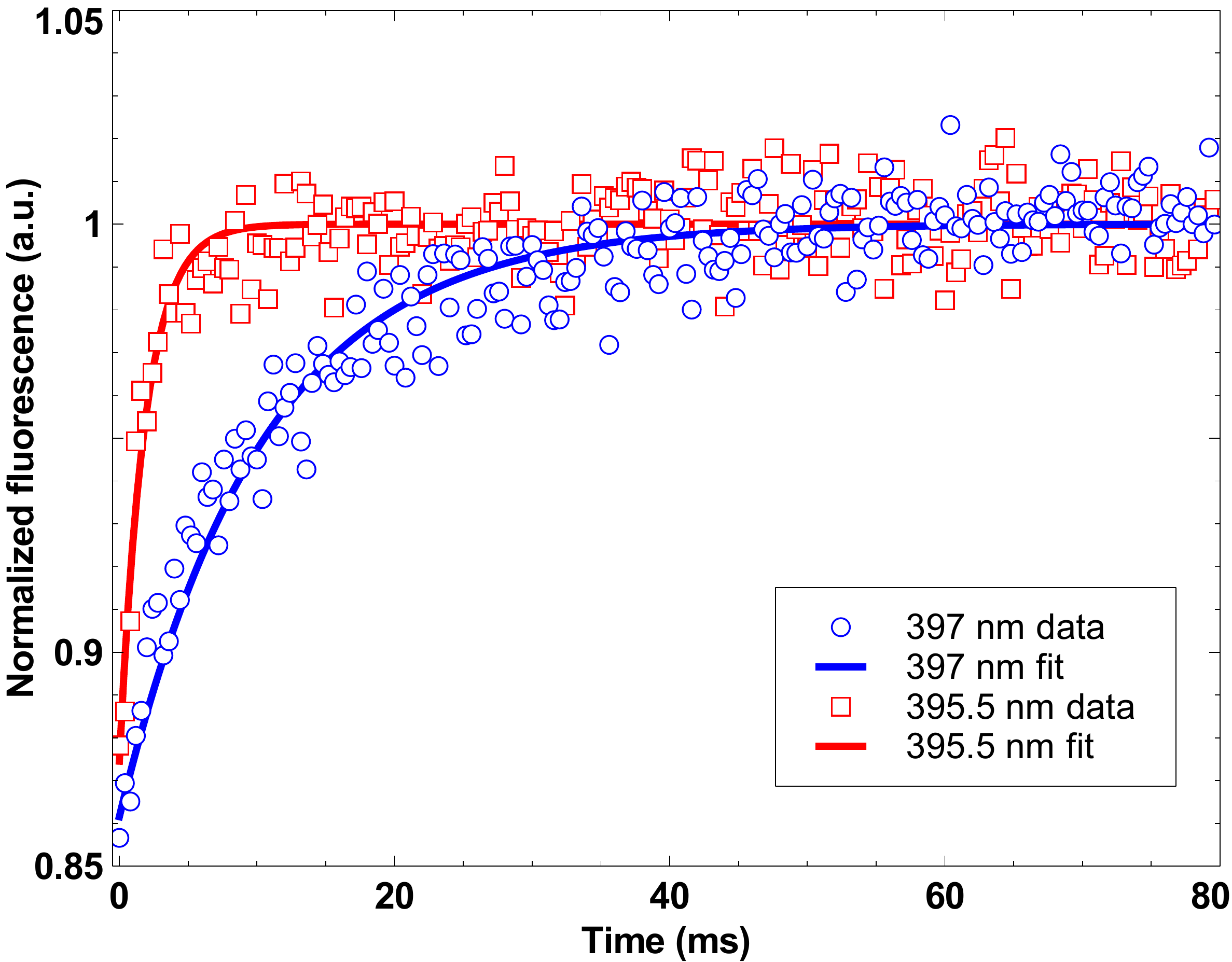}
	\caption{The measured fluorescence recovery curves for excitation at 395.5 nm and 397 nm are shown. The data is fit with a single exponential (Equation \ref{dissociation}) and the fluorescence is normalized to the steady-state fluorescence parameter, $A_{\infty}$, from the fit.} 
	\label{dissociation-plot}
\end{figure*}

We measured the spectrum by repeatedly exciting the ions for 400 $\mu$s before detecting the fluorescence on the PMT for 2 ms as shown in Figure \ref{reaction}. The photon count is recorded as a function of time and the data is fit to a first order reaction rate equation to deduce the dissociation rate:
\begin{equation}\label{dissociation}
A_{t}=A_{\infty}-(A_{\infty}-A_{0})e^{-\Gamma (\lambda) t},
\end{equation}
where $A_{t}$ is the fluorescence at time t, $A_{\infty}$ is the steady-state fluorescence after dissociation, $\Gamma (\lambda)$ is the wavelength dependent rate, and $A_{0}$ is the initial fluorescence. An acoustic-optical modulator (AOM) switches the excitation beam on and off, but the wavelength is scanned manually.  Figure \ref{dissociation-plot} shows example fluoresecence data for two different excitation wavelengths.

\section{Results and discussion}
\begin{figure*}
	\centering
	\includegraphics[width=11.2cm]{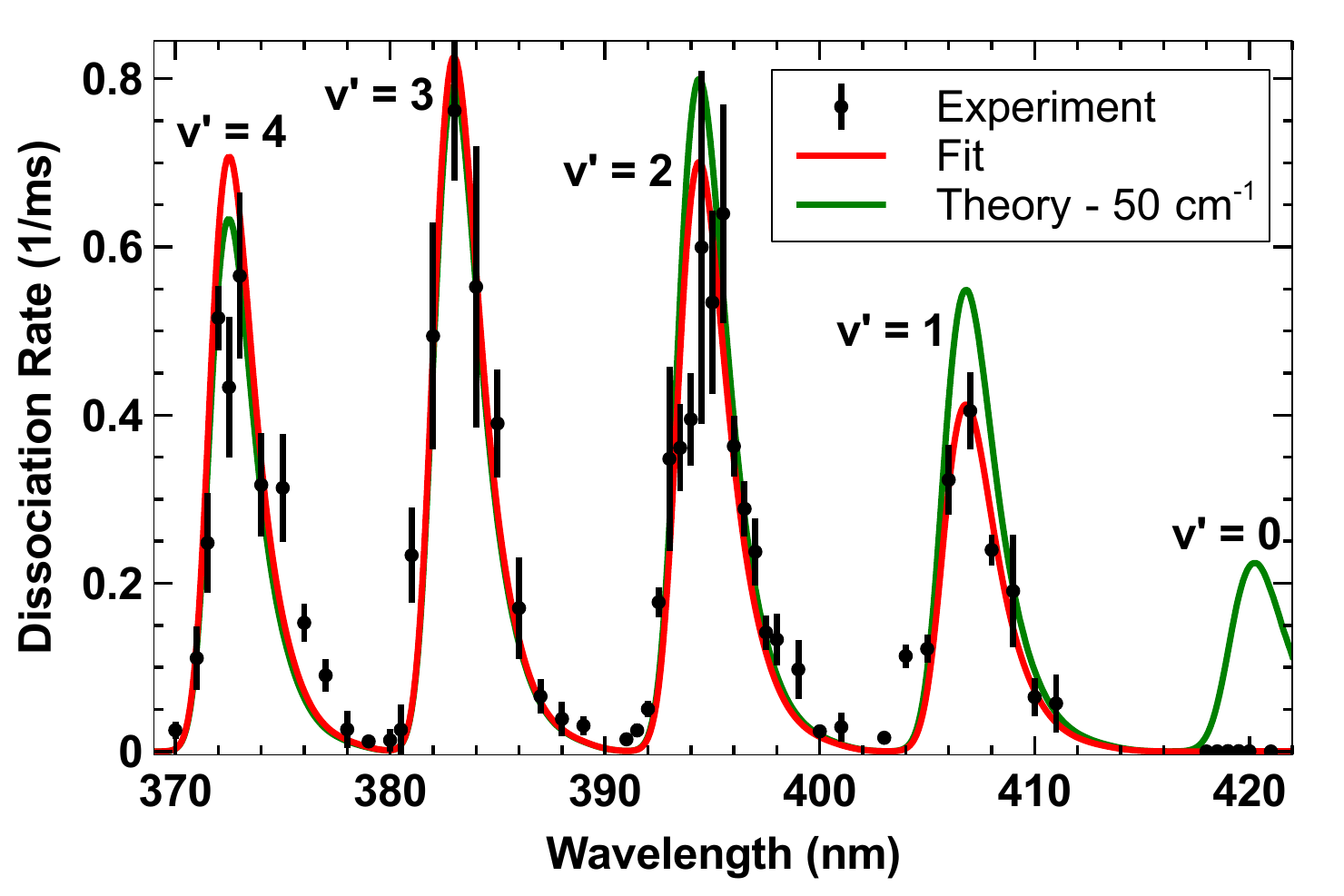}
	\caption{The dissociation rate as a function of excitation wavelength. The error bars are standard deviations over the five measurements. A model that assumes the dissociation rate is limited by the absorption of the first photon is used to plot the theoretical curves and to find the experimental fit. The theory model is based on CASPT2 [31] but all peaks are red shifted by a constant 50 cm$^{-1}$. }
	\label{fig:fitted_experiment}
\end{figure*}

The spectrum acquired is based on the dissociation rate and  we measured the four vibronic transitions shown in Figure \ref{fig:fitted_experiment}. Every single point is an average of five measurements. These rates are repeatable for every wavelength with low standard deviation between experiments as shown in the plot. To avoid variations in raw fluorescence data, we preferred using larger crystals than ion chains with the dissociation beam aligned along the axis of the trap for more homogeneous dissociation across the crystal. Future single-molecule experiments could be performed in the same fashion as in Ref. \cite{ncamiso2015}.

\begin{table*}
	\centering
	\begin{center}
		\begin{tabular}{c c c c}
			\hline
			 $v'$ & \multicolumn{3}{c}{$\nu_{0v'}$}\\
                         & CASPT2 \cite{Abe2011} & CPP-CI \cite{Habli2011} & Experiment\\
			\hline
			$0$ & 23887 & 23826 & (23828) \\ 
			$1$ & 24674 & 24594 & 24624 $\pm$ 13 \\ 
			$2$ & 25449 & 25330 & 25399 $\pm$ 19\\ 
			$3$ & 26206 & 26067 & 26156 $\pm$ 14\\ 
			$4$ & 26942 & 26782 & 26891 $\pm$ 6\\
			\hline
		\end{tabular}
		\caption{Comparison of experimentally measured vibronic transition frequences, 1 $^1\Sigma (v=0) \longrightarrow 2~ ^1\Sigma (v')$, and theoretical calculations. Experimental peaks were assigned based on agreement with theoretical predictions. With this assignment, the transition to $v'=0$ was not measured and the experimental value is extrapolated from the measured peaks ( Equation \ref{anhamonicity-determination})}
		\label{tab:theory compars}
	\end{center}
\end{table*}

We assigned peaks based on theoretical predictions in the literature using a complete-active-space approach (CASPT2)  \cite{Abe2011} and a core pseudopotential - configuration interaction approach (CCP-CI) \cite{Habli2011}. The CASPT2 transition frequencies are blue-shifted and the CCP-PI transition frequencies are red-shifted relative to the data, Table \ref{tab:theory compars}. The CASPT2 predicted transition frequencies differed from the measured transition frequencies by a consistent $\approx$ 50 cm$^{-1}$, which is well within the expected theoretical error based on differences between the calculated asymptotes and the measured atomic spectra of $\approx$ 100 cm$^{-1}$ \cite{matsubara2008}. The difference between the CPP-CI transitions and the measured transitions was less consistent and varied between 30 and 110 cm$^{-1}$. Using this assignment, the  1$^{1}\Sigma, \nu=0 \longrightarrow$ 2$^{1}\Sigma, \nu'=0$ transition was not observed. The experimental peaks showed the expected asymmetry for a system with a smaller rotational constant in the excited state. 

\subsection{Theoretical model for estimation of experimental parameters}

In order to determine the bare vibrational frequency, we modeled our predicted spectra assuming the ion starts in the ground electronic and vibrational state, but a thermal distribution of rotational states.  We truncate the rotational states at $J=15$ which is expected to only have a population of 0.003\% at room temperature. We then calculate the excitation rate using Einstein B coefficients and the spectral density of the doubled Ti:Sapphire. The molecular ion is assumed to instantaneously dissociate after it is excited to the 2 $^1\Sigma$ state.

The rate is calculated as the sum of individual rotational absorption rates multiplied by the thermal probability of being in that state:

\begin{equation}\label{eq:rate}
\Gamma(\lambda) = \sum_{J,v',J'} B_{0,J,v',J'}(\lambda) p(0,J)
\end{equation}
where $B$ is the Einstein stimulated absorption coefficient when the laser is at wavelength $\lambda$ and $p(0,J)$ is the probability of being in the ground vibrational state with the rotational state $J$ assuming a Boltzmann distribution. The Einstein B coefficient is given by:

\begin{equation}
 B_{0,J,v',J'}(\lambda) = \frac{I_{\lambda}(f)}{c} \frac{A_{x,0}}{8 \pi h f^3},
\end{equation}
where $I_{\lambda}(f)$ is the intensity of laser at frequency $F$, $f$ is the transition frequency from 1$^{1}\Sigma, v=0, J \longrightarrow$ 2$^{1}\Sigma, \nu', J'$, and $A_{v',J',0,J}$ is the Einstein A coefficient. Each Einstein A is calculated as:
\begin{equation}
A_{v',J',0,J} = \frac{16 \pi^3 S(J,J')}{3 \varepsilon_0 h} f^3 \mu_{v',0}^2,
\end{equation}
where $J$ is the ground state rotational quantum number, $S(J,J')$ is H\"onl-London factor for a rotational transition $J \longrightarrow J'$, and $\mu$ is the transition dipole moment. 
 
The data for the laser profile was determined by measurement and fitting to a Gaussian profile. The rotational constants for the excited vibronic  levels were calculated using the LEVEL8.2 program \cite{level} and the potential energy surfaces of Ref. \cite{Abe2011}. All other constants were provided by \cite{Abe2011}. Initially, nearly all molecular ions are in the vibronic ground state which means only the $\nu=0$ vibrational state is needed for calculations. 

Figure \ref{fig:fitted_experiment} shows the expected theoretical plots for the CASPT2 with a 50 cm$^{-1}$ red shift. One observes quantitative agreement with the experiemntal peak positions and almost quantitative agreement with the peak heights with the notable exception of the $v'=0$ state.  Our hypothesis is that this is due to the second part of the dissociation process which needs to be further investigated.

To find our experimental fit,  we set the rotational constants to the theoretical values and allow the vibrational transition frequency and dipole matrix element (which controls the peak height) to vary. The nonlinear regression was done by applying Nelder-Mead Minimization Algorithm \cite{NelderMead} to the $\chi^2$ value using a C++ library \cite{ONeill}.  This yields our experimentally determined transition frequencies (Table \ref{tab:theory compars}). 
%The difference between the predicted vibration frequencies from the fit and vibration frequencies based on simply assigning the position to the peak maximum are within  50 cm$^{-1}$

\subsection{Experimental spectroscopic constants}
\begin{figure*}
	\centering
	\includegraphics[width=11.2cm]{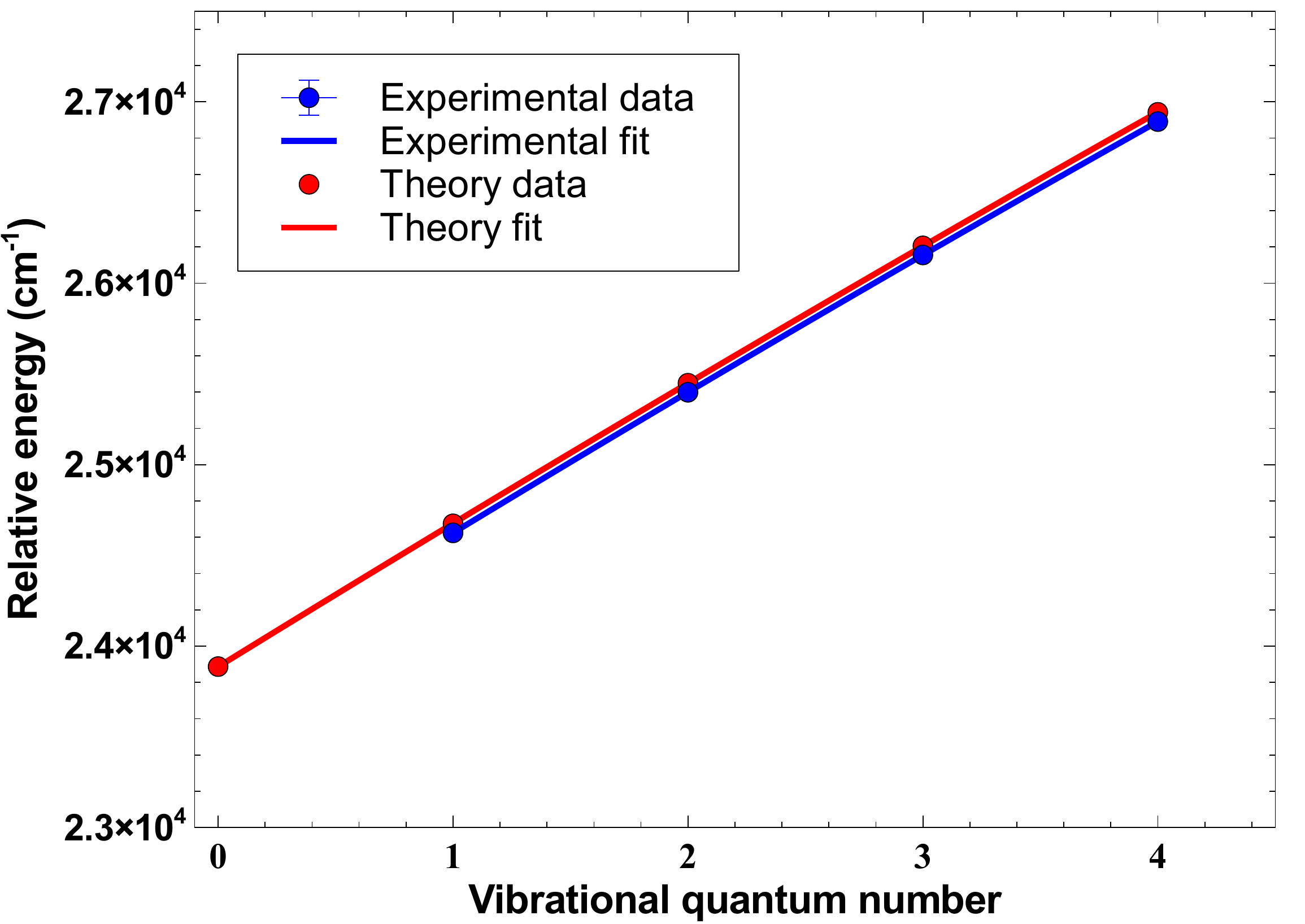}
	\caption{The plot used to determine both the theoretical and experimental values for $\omega_{e}x_{e}$ and $\omega_{e}$ for the the 2$^{1}\Sigma$ state of CaH$^{+}$ is shown. The \textit{ab initio} theoretical energies are taken from ref. \cite{Abe2011}}
	\label{anharmonicity}
\end{figure*}

From our experimental calculation,  we determine the anharmonicity constant ($\omega_{e}x_{e}$) of the molecular ion in the 2$^{1}\Sigma$ state. We fitted both the theoretical vibronic frequencies and our measured spectra to the following equation to determine $\omega_{e}x_{e}$ as shown in Figure \ref{anharmonicity}: 

\begin{equation}\label{anhamonicity-determination}
\nu_{0v'}=\nu_{00}+\omega_{e}v'-\omega_{e}x_{e}v'(v'+1),
\end{equation}
where E(n) is the total vibronic energy, and $\nu_{00}$ is the energy of the $\nu=0 \longrightarrow \nu'=0$  transition; and $\omega_{e}$ is the harmonic constant for the 2$^{1}\Sigma$ state. The theoretical vibronic frequencies used in the fit are from Ref. \cite{Abe2011} . Table \ref{anharmonicity-table} shows our calculated value for the $\omega_{e}x_{e}$ and how it compares to the CASPT2 theoretical results. The fit in Figure \ref{anharmonicity} yielded a value of 10.0 $\pm$  0.2 cm$^{-1}$ for the anharmonicity constant, which is similar to the deduced theoretical prediction of 8.6 $\pm$ 0.3 cm$^{-1}$. The harmonic constant turned out to be 815.8 $\pm$ 1.3 cm$^{-1}$; comparable to the theoretical value of 807 $\pm$ 1.7 cm$^{-1}$. This value also matches the most recent calculations by Aymar and Dulieu \cite{olivier2012}. Furthermore, the fit in Figure \ref{anharmonicity} predicts the $\nu_{00}$ frequency to be 23828 $\pm$ 1.2 cm$^{-1}$ but we did not observe a signal at this frequency.

Our results are a step closer to the the implementation of rotational cooling on CaH$^{+}$ as well as quantum logic spectroscopy on rovibrational transitions in the molecule. 

\begin{table*}
	\begin{center}
		\begin{tabular}{  c c c } 
			\hline
			Constant & Experiment & CASPT2 \cite{Abe2011}\\ 
			\hline
			$\omega_{e}x_{e}$ & 10.0 $\pm$ 0.2 cm$^{-1}$ & 8.6 $\pm$ 0.3 cm$^{-1}$\\ 
			$\omega_{e}$ & 815.8 $\pm$ 1.3 cm$^{-1}$ & 807 $\pm$ 1.8 cm$^{-1}$ \\
			$\nu_{00}$ & 23828 $\pm$ 1.2 cm$^{-1}$ & 23886 $\pm$ 1.2 cm$^{-1}$\\ 
			\hline
		\end{tabular}
		\caption{Table comparing the experimental and theoretical values of the anharmonicity constant as well as the harmonic constant for the 2$^{1}\Sigma$ state of CaH$^{+}$ based on fitting Equation \ref{anhamonicity-determination} to the first five states. The uncertainties are the errors from the fit. We observe good agreement between the experimental and theoretical values for all the constants.} 
		\label{anharmonicity-table}
	\end{center}
\end{table*}

\section{Conclusions}
We have measured the vibronic spectrum of the 1$^{1}\Sigma\longrightarrow$ 2$^{1}\Sigma$  transition in CaH$^{+}$ using resonance enhanced dissociation. The observed peaks were assigned based on previous calculations, which are a reasonable match to the measured results for both the energy spacing between peaks and the expected widths. Our own theoretical modeling of the relative peak intensities agrees with the experimental results with the notable exception of the  $\nu=0 \longrightarrow \nu'=0$ transition. We attribute this discrepancy to the limited experimental characterization of the dissociation process.  

Experimental and theoretical studies on possible dissociation pathways from the 2$^{1}\Sigma$ state are necessary to resolve this conflict. One possibility is that the theoretical calculations are off by 800 cm$^{-1}$, but this seems unlikely for such a simple molecule.  Another solution is the dissociation process involves excitation to the 2$^{1} \Sigma$ state followed by photon emission back to an excited vibrational 1$^{1} \Sigma$ state before photodissociation. Alternatively, the dissociation could be due to predissociation \cite{Christina2002} from the 2$^1\Sigma$ state to the 1$^1\Sigma$ state, but we expect this rate to be slower than fluorescence. In order to unravel these possible pathways, one can look for laser-induced fluourescence when driving these vibronic transitions and also design experiments to measure the electronic state of the dissociated Ca$^+$.

We plan to extend our measurement technique to the measurement of rovibronic lines in CaH$^{+}$ by narrowing the spectrum of our laser source. The measurement of rotational lines is vital for testing the preparation of the molecular ions in their rotational ground state using optical pumping methods. This internal state control, combined with our previous realization of the ground state cooling of axial modes of molecular motion \cite{rugango2015}, will facilitate the implementation of quantum logic spectroscopy on  CaH$^{+}$. In addition, our experimental method can be used to measure the rovibronic spectrum of other alkali-earth monohydride ions with the appropriate laser-cooling ion, as well as the deuterated forms of these molecules.

Regarding state preparation, some optical pumping schemes for rotational cooling will require a good knowledge of the rovibrational spectrum for the ground electronic state. By using a 1+2' resonance enhanced multi-photon dissociation scheme, it is possible to measure the fundamental rovibrational transitions in the 1$^{1}\Sigma$ state by sweeping the frequency of the first photon from a 6.9 $\mu$m laser. The second photon would excite to one of the 2$^{1}\Sigma$ vibrational states, and the third photon would dissociate the molecules \cite{BresselPRL2012, VersolatoPRL2013}. An IR laser source with a wide tuning frequency would enable the measurement of multiple overtones as well.

\section{Acknowledgments}
This work was supported by the Army Research Office (ARO)  (W911NF-12-1-0230), the National Science Foundation (PHY-1404388), and an ARO Multi-University Research Initiative (W911NF-14-1-0378). We would like to thank Minori Abe for providing us with data from \cite{Abe2011}.
%\end{acknowledgments}


\begin{thebibliography}{10}

\bibitem{sandra2014}
S.~Br\"{u}ken, O.~Sipl\"{a}, E.~T. Chambers, J.~Harju, P.~Caselli, O.~Asvany,
  C.~E. Honingh, M.~Kami\'{n}ski, J.~Stutzki, and S.~Schlemmer.
\newblock H$_{2}$\uppercase{D}$^{+}$ observations give an age of at least one
  million years for a cloud core forming sun-like stars.
\newblock {\em Nature}, 56:219, 2014.

\bibitem{maier2015}
E.~K. Campbell, M.~Holz, D.~Gerlich, and J.~P. Maier.
\newblock Laboratory confirmation of \uppercase{C}$_{60}^{+}$ as the carrier of
  two diffuse interstellar bands.
\newblock {\em Nature}, 523:322, 2015.

\bibitem{KajitaJPhysB2011}
M.~Kajita, M.~Abe, M.~Hada, and Y.~Moriwaki.
\newblock Estimated accuracies of pure \uppercase{XH}$^+$ (\uppercase{X}: even
  isotopes of group \uppercase{II} atoms) vibrational transition frequencies:
  towards the test of the variance in m$_p$ / m$_e$.
\newblock {\em J. Phys. B: At. Mol. Opt. Phys.}, 44, 2011.

\bibitem{BresselPRL2012}
U.~Bressel, A.~Borodin, J.~Shen, M.~Hansen, I.~Ernsting, and S.~Schiller.
\newblock Manipulation of individual hyperfine states in cold trapped molecular
  ions and application to \uppercase{HD}$^+$ frequency metrology.
\newblock {\em Phys. Rev. Lett.}, 108, 2012.

\bibitem{LohScience2013}
H.~Loh, K.~C. Cossel, M.~C. Grau, K.-K. Ni, E.~R. Meyer, J.~L. Bohn, J.~Ye, and
  E.~A. Cornell.
\newblock Precision spectroscopy of polarized molecules in an ion trap.
\newblock {\em Science}, 342:1220--1222, 2013.

\bibitem{rugango2015}
R.~Rugango, J.~E. Goeders, T.~H. Dixon, J.~M. Gray, N.~Khanyile, G.~Shu, R.~J.
  Clark, and K.~R. Brown.
\newblock Sympathetic cooling of molecular ion motion to the ground state.
\newblock {\em New. J. Phys.}, 17:035009, 2015.

\bibitem{wan2015}
Y.~Wan, F.~Wolf, F.~Gebert, and P.~O. Schmidt.
\newblock Efficient sympathetic motional ground-state cooling of a molecular
  ion.
\newblock {\em Phys. Rev. A}, 91:043425, 2015.

\bibitem{LienNatComm2014}
{C.-Y.} Lien, C.M. Seck, {Y.-W.} Lin, J.H.V. Nguyen, D.~A. Tabor, and B.~C.
  Odom.
\newblock Broadband optical cooling of molecular rotors from room temperature
  to the ground state.
\newblock {\em Nat. Commun.}, 5:4783, 2014.

\bibitem{StaanumNatPhys2010}
P.~F. Staanum, K.~H{\o}jbjerre, P.~S. Skyt, A.~K. Hansen, and M.~Drewsen.
\newblock {Rotational laser cooling of vibrationally and translationally cold
  molecular ions}.
\newblock {\em Nat. Phys.}, 6:271, 2010.

\bibitem{SchneiderNatPhys2010}
T.~Schneider, B.~Roth, H.~Duncker, I.~Ernsting, and S.~Schiller.
\newblock {All-optical preparation of molecular ions in the rovibrational
  ground state}.
\newblock {\em Nat. Phys.}, 6:275, 2010.

\bibitem{wolf2015}
F.~Wolf, Y.~Wan, J.~C. Heip, F.~Gebert, C.~Shi, and P.~O. Schmidt.
\newblock Non-destructive state detection for quantum logic spectroscopy of
  molecular ions.
\newblock {\em Nature}, 530:457, 2016.

\bibitem{HansenNature2014}
A.~K. Hansen, O.~O. Versolato, L.~Klosowski, A.~Gingell, M.~Schwarz,
  A.~Windberger, J.~Ullrich, L.~Crespo, and M.~Drewsen.
\newblock Efficient rotational cooling of coulomb-crystallized molecular ions
  by a helium buffer gas.
\newblock {\em Nature}, 508:76, 2014.

\bibitem{Schiller2014}
S.~Schiller, D.~Bakalov, A.~K. Bekbaev, and V.~I. Korobov.
\newblock Static and dynamic polarizability and the stark and
  blackbody-radiation frequency shifts of the molecular hydrogen ions
  \uppercase{H}$_{2}^{+}$, \uppercase{HD}$^+$, and \uppercase{D}$_{2}^{+}$.
\newblock {\em Phys. Rev. A}, 89:052521, 2014.

\bibitem{Germann2014}
M.~Germann, X.~Tong, and S.~Willitsch.
\newblock Observation of electric-dipole-forbidden infrared transitions in cold
  molecular ions.
\newblock {\em Nat. Phys.}, 10:820--824, 2014.

\bibitem{KajitaPRA2014}
M.~Kajita, G.~Gopakumar, M.~Abe, M.~Hada, and M.~Keller.
\newblock Test of ${m}_{p}$/${m}_{e}$ changes using vibrational transitions in
  {N}${}_{2}{}^{+}$.
\newblock {\em Phys. Rev. A}, 89:032509, Mar 2014.

\bibitem{albert1909}
A.~Eagle.
\newblock On the spectra of some of the compounds of the alkaline earths.
\newblock {\em Astrophys. J.}, 30:231, 1909.

\bibitem{mcall2011}
A.~A. Mills, B.~M. Siller, M.~W. Porambo, M.~Perera, H.~Kreckel, and B.~J.
  McCall.
\newblock Ultra-sensitive high-precision spectroscopy of a fast molecular ion
  beam.
\newblock {\em J. Chem. Phys.}, 135:224201, 2011.

\bibitem{kimura}
N.~Kimura, K.~Okada, T.~Takayanagi, M.~Wada, S.~Ohtani, and H.~A. Schuessler.
\newblock Sympathetic crystallization of {CaH}$^{+}$ produced by a
  laser-induced reaction.
\newblock {\em Phys. Rev. A}, 83:033422, 2011.

\bibitem{schiller2006}
B.~Roth, H.~Wenz, H.~Daerr, and S.~Schiller.
\newblock Ion-neutral chemical reactions between ultracold localized ions and
  neutral molecules with single-particle resolution.
\newblock {\em Phys. Rev. A}, 73:042712, 2006.

\bibitem{MolhavePRA2000}
K.~M{\o}lhave and M.~Drewsen.
\newblock {Formation of translationally cold MgH$^{+}$ and MgD$^{+}$ molecules
  in an ion trap}.
\newblock {\em Phys. Rev. A}, 62:011401, 2000.

\bibitem{HansenAngew2012}
A.~K. Hansen, M.~A. S{\o}rensen, P.~F. Staanum, and M.~Drewsen.
\newblock Single-ion recycling reactions.
\newblock {\em Angew. Chem. Int. Ed.}, 51:7960, 2012.

\bibitem{ncamiso2015}
N.~B. Khanyile, G.~Shu, and K.~R. Brown.
\newblock Observation of vibrational overtones by single-molecule resonant
  photodissociation.
\newblock {\em Nature Commun.}, 6:7825, 2015.

\bibitem{Flohlich2004}
U.~Fl\"{o}hlich, B.~Roth, P.~Antonini, C.~L. Ammerzahl, A.~Wicht, and
  S.~Schiller.
\newblock Novel systems for tests of the time-independence of the
  electron-to-proton mass ratio.
\newblock {\em Lect. Notes Phys.}, 648:297, 2004.

\bibitem{Roth2005}
B.~Roth, J.~Koelemeij, S.~Schiller, L.~Hilico, and J.~P. Karr.
\newblock Precision spectroscopy of molecular hydrogen ions: Towards frequency
  metrology of particle masses.
\newblock {\em Lect. Notes Phys.}, 745:205, 2008.

\bibitem{krems}
R.~Krems, F.~Bretonisch, and W.~C. Stwalley.
\newblock {\em Cold Molecules: Theory, Experiment, Applications}.
\newblock CRC Press, 2009.

\bibitem{kajita2012}
M.~Kajita and M.~Abe.
\newblock Frequency uncertainty estimation for the
  $^{40}$\uppercase{C}a\uppercase{H}$^{+}$ vibrational transition frequencies
  observed by \uppercase{R}aman excitation.
\newblock {\em J. Phys. B: At. Mol. Opt. Phys.}, 45:185401, 2012.

\bibitem{AbeKajita}
M.~Abe, M.~Kajita, M.~Hada, and Y.~Moriwaki.
\newblock Ab initio study on vibrational dipole moments of \uppercase{XH}$^+$
  molecular ions: \uppercase{X} = $^{24}$\uppercase{M}g, $^{40}$\uppercase{C}a,
  $^{64}$\uppercase{Z}n, $^{88}$\uppercase{S}r, $^{114}$\uppercase{C}d,
  $^{138}$\uppercase{B}a, $^{174}$\uppercase{Y}b, and $^{202}$\uppercase{H}g.
\newblock {\em J. Phys. B: At. Mol. Opt. Phys.}, 43:245102, 2010.

\bibitem{olivier2012}
M.~Aymar and O.~Dulieu.
\newblock The electronic structure of the alkaline-earth-atom (\uppercase{C}a,
  \uppercase{S}r, \uppercase{B}a) hydride molecular ions.
\newblock {\em J. Phys. B: At. Mol. Opt. Phys.}, 45:215103, 2012.

\bibitem{goeders}
J.~E. Goeders, C.~R. Clark, G.~Vittorini, K.~Wright, R.~C. Viteri, and K.~R.
  Brown.
\newblock Identifying single molecular ions by resolved sideband measurements.
\newblock {\em J. Phys. Chem. A}, 117:9725, 2013.

\bibitem{Abe2011}
M.~Abe, Y.~Moriwaki, M.~Hada, and M.~Kajita.
\newblock Ab initio study on potential energy curves of electronic ground and
  excited states of $^{40}$\uppercase{C}a\uppercase{H}$^+$ molecule.
\newblock {\em Chem. Phys. Lett.}, 521:31, 2011.

\bibitem{Habli2011}
H.~Habli, H.~Ghalla, B.~Oujia, and F.X. Gad{\'e}a.
\newblock Ab initio study of spectroscopic properties of the calcium hydride
  molecular ion.
\newblock {\em Eur. Phys. J. D}, 64:5, 2011.

\bibitem{matsubara2008}
K.~Matsubara, K.~Hayasaka, Y.~Li, Y.~Ito, S.~Nagano, M.~Kajita, and M~Hosokawa.
\newblock Frequency measurement of the optical clock transition of
  $^{40}$\uppercase{C}a$^{+}$ ions with an uncertainty of 10$^{-14}$ level.
\newblock {\em App. Phy. Exp.}, 1:067011, 2008.

\bibitem{level}
R.~J.~Le Roy.
\newblock Level 8.2: A computer program for solving the radial schr\"{o}dinger
  equation for bound and quasibound levels.
\newblock {\em University of Waterloo Chemical Physics Research Report},
  CP-663, 2014.

\bibitem{NelderMead}
J.~Nelder and R.~Mead.
\newblock A simplex method for function minimization.
\newblock {\em Comput. J.}, 7:308, 1965.

\bibitem{ONeill}
R.~ONeill.
\newblock Algorithm as 47: Function minimization using a simplex procedure.
\newblock {\em J. Appl. Stat.}, 20:338, 1971.

\bibitem{Christina2002}
C.~Carlsund-Levin, N.~Elander, A.~Nunez, and A.~Scrinzi.
\newblock An exterior complex rotated coupled channel description of
  predissociationin diatomic molecules applied to a model of the four lowest
  $^{2}\sigma^{+}$-states in cah.
\newblock {\em Physica Scripta}, 65:306, 2002.

\bibitem{VersolatoPRL2013}
O.~O. Versolato, M.~Schwarz, A.~K. Hansen, A.~D. Gingell, A.~Windberger, \L{}.
  Klosowski, J.~Ullrich, F.~Jensen, L.~Crespo, and M.~Drewsen.
\newblock Decay rate measurement of the first vibrationally excited state of
  \uppercase{M}g\uppercase{H}$^+$ in a cryogenic \uppercase{P}aul trap.
\newblock {\em Phys. Rev. Lett.}, 111:053002, 2013.

\end{thebibliography}
\end{document}